\documentclass[aip,jap,reprint]{revtex4-1}

\usepackage{graphicx}
\usepackage{amsmath}
\usepackage{siunitx}
\usepackage{sidecap}
\usepackage{pdfpages}

\begin{document}


\title{Non-ferroelectric nature of the conductance hysteresis in CH$_3$NH$_3$PbI$_3$ perovskite-based photovoltaic devices}

\author{J. Beilsten-Edmands}
\affiliation{Department of Physics, Clarendon Laboratory, University of Oxford, Oxford OX1 3PU, United Kingdom}
\email{james.beilsten-edmands@physics.ox.ac.uk}
\author{G. E. Eperon}
\affiliation{Department of Physics, Clarendon Laboratory, University of Oxford, Oxford OX1 3PU, United Kingdom}
\author{R. D. Johnson}
\affiliation{Department of Physics, Clarendon Laboratory, University of Oxford, Oxford OX1 3PU, United Kingdom}
\author{H. J. Snaith}
\affiliation{Department of Physics, Clarendon Laboratory, University of Oxford, Oxford OX1 3PU, United Kingdom}
\author{P. G. Radaelli}
\affiliation{Department of Physics, Clarendon Laboratory, University of Oxford, Oxford OX1 3PU, United Kingdom}





\begin{abstract}
We present measurements of conductance hysteresis on CH$_3$NH$_3$PbI$_3$ perovskite thin films, performed using the double-wave method, in order to investigate the possibility of a ferroelectric response.  A strong frequency dependence of the hysteresis is observed in the range of \SI{0.1}{\hertz} to \SI{150}{\hertz}, with a hysteretic charge density in excess of \SI{1000}{\micro\coulomb\per\square\centi\metre} at frequencies below \SI{0.4}{\hertz} --- a behaviour uncharacteristic of a ferroelectric response.  We show that the observed hysteretic conductance, as well as the presence of a double arc in the impedance spectroscopy, can be fully explained by the migration of mobile ions under bias on a timescale of seconds. Our measurements place an upper limit of $\approx \SI{1}{\micro\coulomb\per\square\centi\metre}$ on any intrinsic frequency-independent polarisation, ruling out ferroelectricity as the main cause of current-voltage hysteresis and providing further evidence of the importance of ionic migration in modifying the efficiency of CH$_3$NH$_3$PbI$_3$ devices.

\end{abstract}

     
\maketitle

Solar cells based on a layer of halide perovskite as the active material have recently seen rapid increases in efficiency, with certified values in excess of 20\% \cite{NREL}. In these devices, a hysteresis effect has been observed in the current-voltage (JV) curves, whereby the photocurrent depends on the direction of the voltage sweep between forward bias and short circuit configurations. Understanding this hysteresis effect is important for characterising the steady-state efficiency of such devices reliably under working conditions \cite{Gratzel}.
Hysteresis has been reported to depend on scan rate, contact material \cite{Snaith14}, biasing history \cite{Miyasaka}, and lighting conditions \cite{Unger}. In addition, time dependent photoconductivity responses have also been observed \cite{Snaith14, Gottesman}.
Several explanations for these effects have been proposed \cite{Snaith14}, including charge traps at the perovskite surface,  ion migration\cite{Unger} leading to band-bending at the interface \cite{HuangNatMat, Tress}, and the intrinsic ferroelectric nature of the perovskite material \cite{Ono, Frost14, Wei14, Miyasaka, Unger}. 

The possibility of ferroelectricity has been suggested since, in general, the perovskite crystal structure is known to undergo a range of distortions, including polar distortions, thereby allowing a ferroelectric polarisation to develop. This theory has gained momentum after a recent crystallographic study of the room temperature tetragonal structure reported the polar $\it{I}$4$\it{cm}$ space group \cite{Kanatzidis}, rather than the previously accepted non-polar $\it{I}$4/$\it{mcm}$  \cite{Poglitsch, Kawamura02, Weller}. In addition to the potential polarisability of the Pb-I lattice, orientational ordering of the methylammonium (MA) molecular dipoles could lead to molecular ferroelectricity or antiferroelectricity \cite{Frost14}. 
 \textit{Ab initio} calculations have predicted a ferroelectric polarisation (P) of \SI{38}{\micro\coulomb\per\square\centi\metre}  for CH$_3$NH$_3$PbI$_3$ \cite{Frost14}, while measurements indicating a ferroelectric response have been reported in the literature\cite{Kanatzidis, Wei14, Kutes, BChen}. However, there has yet to be any conclusive evidence of the ferroelectric properties of this material. 
Although not intrinsically hysteretic, dielectric effects must also be considered, as an applied voltage will induce a dielectric polarisation, which may affect device operation. A dielectric study on CH$_3$NH$_3$PbI$_3$ revealed a picosecond relaxation process, corresponding to dynamic disorder of the MA ions \cite{Poglitsch}, whereas other studies have identified a low frequency response, also attributed to dielectric relaxation \cite{Sanchez, Pascoe}.

We set out to investigate whether ferroelectricity is the cause of the observed JV hysteresis in CH$_3$NH$_3$PbI$_3$ thin films. 
Using the double-wave method (DWM)\cite{Fukunaga08}, also known as the PUND method\cite{PUND}, we observed polarisation-electric field (PE) hysteresis loops, in agreement with previous studies \cite{Kanatzidis, Wei14, BChen}.  However, the giant low-frequency magnitude of the hysteretic charge density and its large frequency dependence are clear demonstrations that this response cannot be due to ferroelectricity. Based on the observed charging processes, we attribute the PE hysteresis to ionic migration under applied voltage and the band-bending that this induces, and we show how this process manifests as JV hysteresis. Therefore, we believe that most attempted measurements of ferroelectricity reported so far have wrongly labelled the effect observed, and that this term should not be applied to the switchable behaviour in methylammonium halide perovskites.

\begin{figure}
\includegraphics[width=0.5\textwidth]{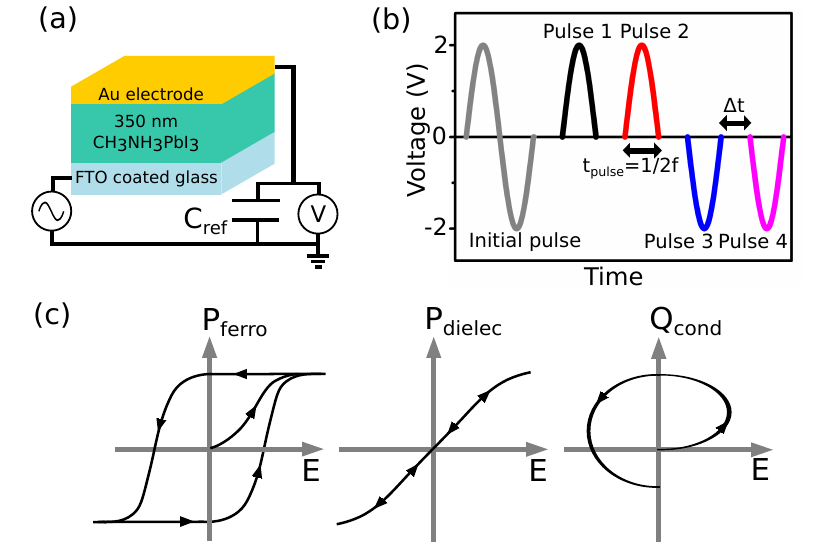}
\caption{(a) Experimental circuit and (b) applied waveform used for the DWM. A voltage amplitude of \SI{2}{\volt} was used, corresponding to an E-field of \SI{5.7}{\kilo\volt\per\milli\metre}. Measurements were made across a frequency range of \SI{150}{\hertz} to \SI{0.1}{\hertz}, corresponding to pulse width times $t_{\text{pulse}}$ from \SI{1/300}{\second} to \SI{5}{\second}. Gap times $\Delta t$ between \SI{5}{\second} and \SI{10}{\second} were necessary to discharge C$_{\text{ref}}$.  
(c) Expected E-field dependence of accumulated charges due to ferroelectric polarisation, dielectric polarisation and diode conduction. For semiconductive materials $Q_{\text{cond}}$ can be orders of magnitude larger than $Q_{\text{Ferro}}$ = $P_{\text{Ferro}}\times A$, where $A$ is the contact area.}
\label{fig:experimental}
\end{figure} 

\begin{SCfigure*}
\includegraphics[width=0.65\textwidth]{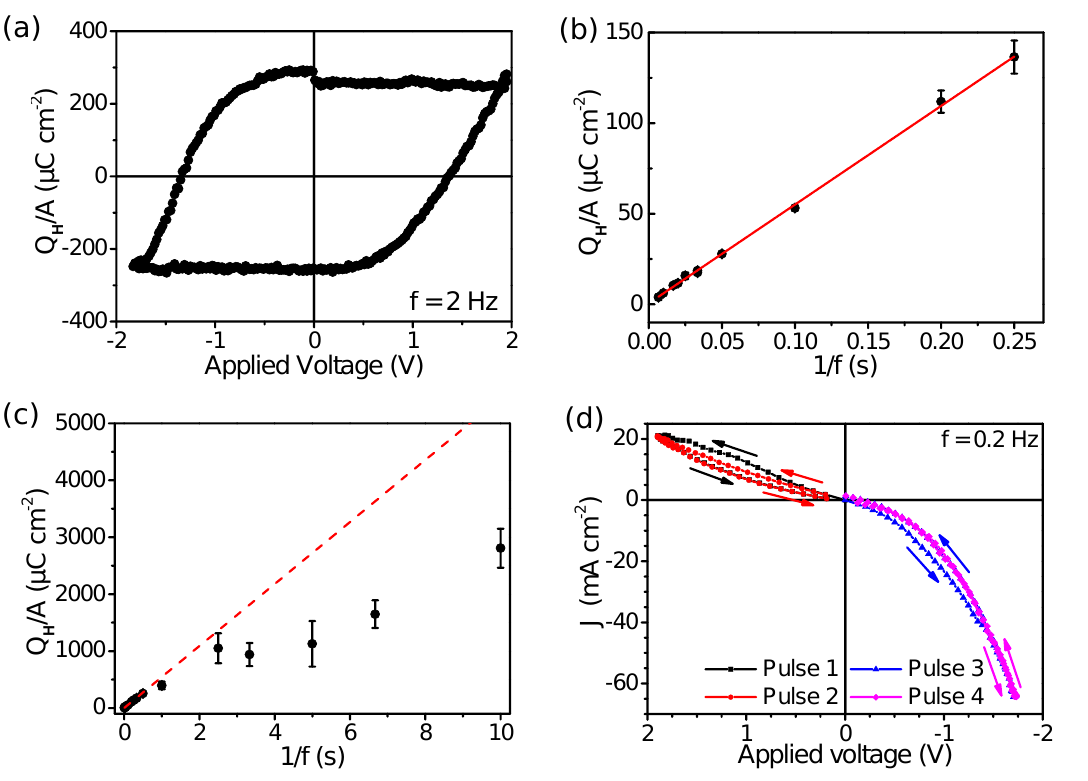}
\caption{(a) Hysteretic charge density (Q$_{\text{H}}$/A) as measured by the DWM on a CH$_3$NH$_3$PbI$_3$ thin film at \SI{2}{\hertz}. (b) Frequency dependence of the QV hysteresis as a function of $1/f$ below \SI{4}{\hertz} (circles). The continuous line is a linear fit to the data for $1/f \leq 0.25$. (c) Frequency dependence of the QV hysteresis at low frequencies, where some saturation from the low frequency linear behaviour (dotted line) is observed. For $f$ below \SI{1}{\hertz}, error bars indicate the range of measured values of hysteresis at each frequency. (d) DWM measurement plotted as JV curves. Arrows indicate voltage sweep direction. The QV hysteresis arises from a difference in current density between pulses 1 \& 2, and pulses 3 \& 4. The voltage axis has been reversed to display the JV curve in the orientation typically shown in solar cell literature.}
\label{fig:Fdep}
\end{SCfigure*}

The samples for hysteresis measurements were pinhole-free \SI{350}{\nano\metre} thick CH$_3$NH$_3$PbI$_3$ thin films sandwiched between two conductive electrodes. The CH$_3$NH$_3$PbI$_3$, fabricated using a technique previously described as vapour-assisted solution processing \cite{ChenQi}, was contacted on one side with fluorine-doped tin oxide and on the other with a gold electrode. Full details of fabrication and growth are described in the supplementary information \cite{supplemental}. 
The potential ferroelectric response was investigated using the DWM, in order to separate out unambiguously hysteretic behaviour from conductive and dielectric effects.
The methodology of the DWM is fully described in Ref. \citenum{Fukunaga08}; here we provide a brief explanation of our measurement. A voltage pulse was applied across the film, which was in series with a large reference capacitor C$_{\text{ref}}$ as shown in Figure~\ref{fig:experimental}(a). The capacitor integrates the surface charge due to ferroelectric and dielectric polarisation ($Q$ = $P \times A$, where $A$ is the contact area) plus the conductive current of the film during the measurement. 
C$_{\text{ref}}$ was chosen to fulfil the critereon $R_{\text{sample}}$C$_{\text{ref}}$ $\gg$ $t_{pulse}$ such that additional transient behaviour was not introduced (with values ranging from \SI{2.2}{\milli\farad} at \SI{150}{\hertz} to \SI{150}{\milli\farad} at \SI{0.1}{\hertz}). This maintained the capacitor voltage below \SI{0.2}{\volt} and this back-voltage was corrected for in the loop analysis. The applied voltage waveform is shown in Figure~\ref{fig:experimental}(b) and experimental parameters are given in the figure caption. The waveform consisted of a sine wave of one period, followed by two half period sine waves in the positive direction and two in the negative direction, hereafter referred to as the initial pulse and pulses 1 to 4. Gap times ranging between 5 and 10 seconds were necessary to discharge C$_{\text{ref}}$ between pulses, to avoid the introduction of additional hysteretic behaviour. The initial pulse would pole a ferroelectric film in the `down' state. The integrated charge from pulse 1 would contain any hysteretic switching behaviour, as well as additional contributions from non-hysteretic effects such as conduction and dielectric polarisation. Figure~\ref{fig:experimental}(c) shows the expected electric-field dependence of these contributions. As the film is now in the `up' state, pulse 2 gives only the unwanted additional contributions, and subtraction yields the underlying hysteretic behaviour. This process is repeated for the negative direction, enabling the full charge-voltage (QV) hysteresis loop (which can be expressed as a PE hysteresis loop) to be analytically reconstructed. 
All measurements were made under dark conditions, to eliminate the photocurrent and measure the intrinsic characteristics of the films.

Figure~\ref{fig:Fdep}(a) shows a hysteresis loop obtained by the DWM on a CH$_3$NH$_3$PbI$_3$ thin film.
The hysteresis is plotted in units of \SI{}{\micro\coulomb\per\square\centi\metre}, in order to compare to known ferroelectrics. 
Figures~\ref{fig:Fdep}(b),(c) show the frequency dependence of the hysteretic charge density measured under positive bias.
Remarkably, the shape of the hysteresis loop looks very similar to the one expected for a ferroelectric material but the \emph{magnitude} of the hysteretic charge density exceeds \SI{1000}{\micro\coulomb\per\square\centi\metre} below \SI{0.4}{\hertz} --- over an order of magnitude larger than for perovskite ferroelectrics. For example, the polarisation of bulk BaTiO$_3$ is \SI{15}{\micro\coulomb\per\square\centi\metre} (Ref. \citenum{Hippel50}), and is enhanced to \SI{70}{\micro\coulomb\per\square\centi\metre} in thin films \cite{Choi05}, while the largest reported polarisation is on the order of \SI{150}{\micro\coulomb\per\square\centi\metre} for BiFeO$_3$ thin films \cite{Kwi}. Hence, our measured hysteresis in CH$_3$NH$_3$PbI$_3$ is very unlikely to arise from ferroelectricity. 
As explained previously, this large charge density cannot simply be attributed to ferroelectricity plus an additional contribution due to electronic conduction, as this is corrected for by the DWM.
Additionally, between \SI{150}{\hertz} and \SI{4}{\hertz}, the measured hysteresis is to a good approximation linear in $1/f$ (i.e. linear in the time of applied voltage). This is uncharacteristic of the switching kinetics observed in ferroelectric thin films \cite{Tagantsev}, where ferroelectric polarisation is expected to be frequency independent to good approximation, particularly in the low frequency regime.  The giant magnitude of the hysteretic charge density requires a dominant non-ferroelectric process, which, as explained in the remainder of the paper, we identify as due to ion migration and trapping at the interfaces.  Since these effects are minimised at high frequency, we can define an upper limit to the `true' ferroelectric polarisation as the high-frequency intercept of the $Q_{\text{H}}$ vs $1/f$ line.  As $1/f \rightarrow 0$, the linear fit was found to pass through the origin within error, and the uncertainty puts an upper limit on any intrinsic frequency-independent polarisation of \SI{1}{\micro\coulomb\per\square\centi\metre}. Although this uncertainty is not insignificant, we can state with confidence that our measurements rule out a large polarisation as predicted by Frost \textit{et al.} \cite{Frost14}. It should be noted that if there were a ferroelectric molecular dipole ordering that had a slow response to an external field (i.e. at low frequencies), this would not be visible in our measurements due to the dominant non-ferroelectric contribution to hysteresis at low frequency. 
The absence of ferroelectricity is in agreement with the majority of reports in the literature on the crystallography of CH$_3$NH$_3$PbI$_3$, which determine a non-ferroelectric structure described by the $\it{I}$4/$\it{mcm}$ space group \cite{Poglitsch, Kawamura02, Weller}.  A polar distortion of the Pb-I lattice or ferroelectric ordering of MA dipoles would necessarily lower the symmetry of the whole crystal to a polar space group.

In order to establish the true origin of the QV hysteresis, the majority of the frequency dependence can be removed by considering the current density, rather than accumulated charge. Figure~\ref{fig:Fdep}(d) shows the dark JV curve as measured by the DWM at \SI{0.2}{\hertz}. The measured hysteresis is now visible as a difference between the current density extracted from the film on the first and second (or third and fourth) pulses, with this JV hysteresis being frequency independent above \SI{1}{\hertz}, and saturating to some extent below \SI{1}{\hertz}. It is this frequency independent JV hysteresis above \SI{1}{\hertz} that gives rise to a QV hysteresis linear in $1/f$ when the current is integrated in the DWM measurement. A reduction in absolute current density as the frequency was reduced was also observed.

\begin{SCfigure*}
\includegraphics[width=0.7\textwidth]{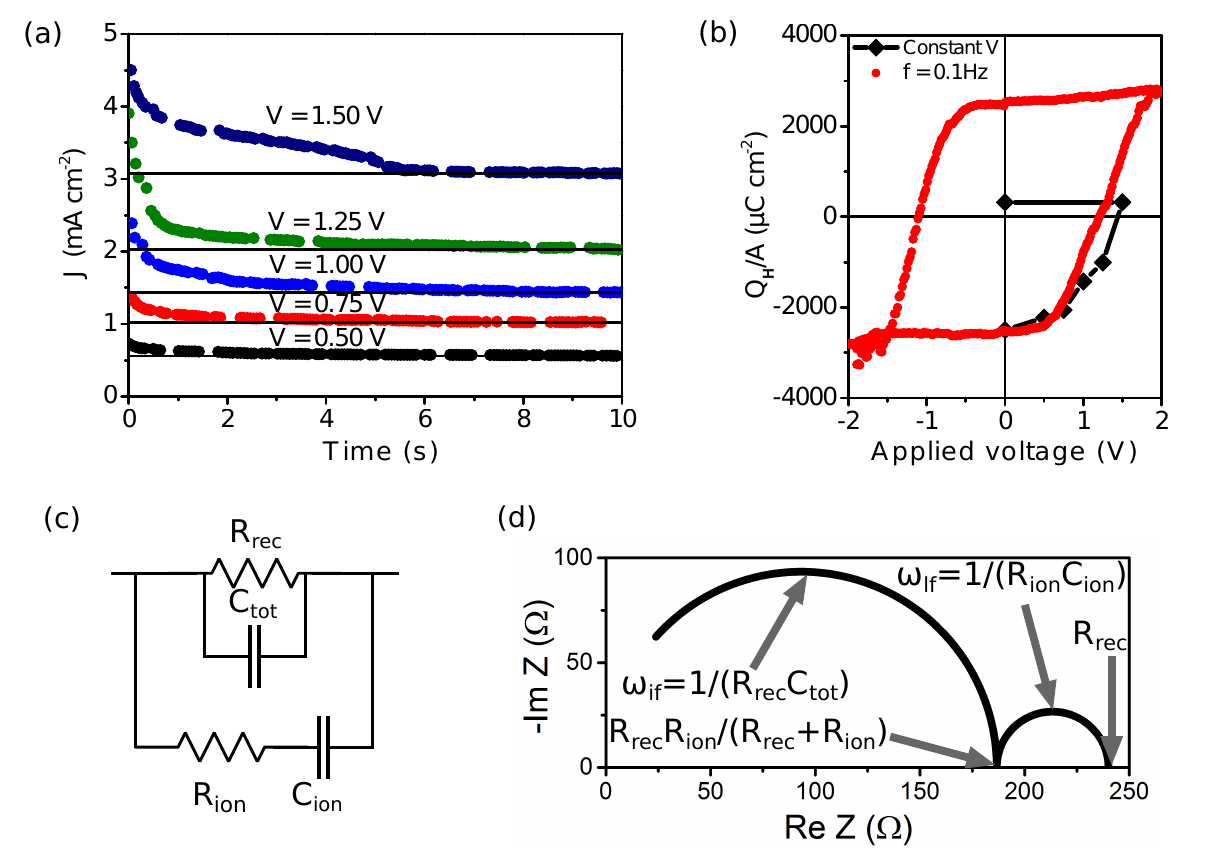}
\caption{(a) Time-dependent current density (circles) at a series of fixed voltages V, after poling at -V.  The solid lines indicate the saturated steady-state backgrounds. (b) Comparison of the integrated additional charge at fixed V and DWM measurement at \SI{0.1}{\hertz}. The integrated charge is shifted by \SI{-2500}{\micro\coulomb\per\square\centi\metre} such that both curves have a common origin at (0,-2500). (c) Equivalent circuit for our thin films, including the effect of ionic migration. (d) Calculated impedance response of the equivalent circuit, using $C_{\text{tot}}=\SI{14}{\nano\farad}$ and the experimentally determined parameters $R_{\text{rec}} =\SI{240}{\ohm}$, $R_{\text{ion}}=\SI{840}{\ohm}$ and $C_{\text{ion}}=\SI{0.28}{\milli\farad}$. The existence of the ionic component in the equivalent circuit produces an additional arc at low frequencies.}
\label{fig:3}
\end{SCfigure*}

To further investigate the origin of JV hysteresis, we measured the time-dependent response from the film at fixed applied voltage V, after initial poling of -V, representative of DC switching.
Figure~\ref{fig:3}(a) shows the time-dependent current density at a series of applied voltages, measured on a single film. 
We note that the results measured on this film, discussed in the following, were fully representative of a number of films measured.
It was not possible to apply constant voltages above \SI{1.5}{\volt} as this was found to damage the films. In the constant voltage measurements we observed a lower overall current density compared to the DWM measurements, which we attribute to film degradation over time.  
At fixed voltage, the initially observed current is greater than the steady-state current, which is reached after approximately 10 seconds. This is indicative of a charge-trapping process at the perovskite/electrode interface that saturates on the timescale of seconds. 
Although this could in principle be due to either electronic trapping or ionic trapping, given the high electronic mobility of this compound \cite{Kanatzidis}, one would expect electronic trapping on the order of nanoseconds. Therefore this charge should be attributed to ionic migration, which has indeed been previously reported to occur on the timescale of seconds in CH$_3$NH$_3$PbI$_3$ devices \cite{HuangNatMat}. Perovskite halides are known to be ionic conductors \cite{Mizusaki, Kuku} and the applied maximum field of \SI{5.7}{\kilo\volt\per\milli\metre} should induce significant migration.  Upon removing the voltage, the discharging current was negligible, showing that the additional charge had indeed been trapped.  
To compare directly to our JV and DWM measurements, we integrated the additional charge above the steady state background at \SI{10}{\second} at each fixed voltage, between 0 and \SI{10}{\second}. Figure~\ref{fig:3}(b) shows a comparison of the lowest frequency DWM hysteresis loop and a hysteresis loop constructed from the integrated excess charge after switching at constant voltage. The reduced magnitude of accumulated charge in the constant voltage measurement is attributed to current-induced fatigue, as previously stated. The agreement between the two methods reveals how ferroelectric-like hysteresis loops can result from slowly saturating ionic charges and ionic trapping at the interfaces, where the role of the coercive electric field is played by the activation energy for ionic de-trapping.  Due to the time dependent nature of the voltage waveform in the DWM, at later times the ionic current is always reduced for a given voltage, resulting in a difference once the pulses are subtracted.  
We remark that, in addition to charge trapping, any change in the carrier extraction efficiency at the perovskite/electrode interface due to migration-induced band-bending  \cite{HuangNatMat} will modify the steady-state background and hence will also give a contribution to the additional charge observed, in both sets of measurements. Our measurements also enable us to place an upper limit on the fraction of ions required to give such an accumulated charge.   As an example, taking the $\text{I}^{-}$ ions to be the ionic species, the hysteresis of \SI{2500}{\micro\coulomb\per\square\centi\metre} measured using the DWM corresponds to an upper limit of 3.7\% of the $\text{I}^{-}$ ions present in the film. In reality it is likely that all species give some contribution to ionic migration since, for CH$_3$NH$_3$PbI$_3$, vacancies of MA, Pb and I have all been calculated to have low formation energies (below \SI{2}{eV}), depending on the chemical potential \cite{Yin14}. 
It is important to note that whilst our DWM measurements start at \SI{0}{\volt}, typically JV measurements in the literature are started from open circuit, equivalent to measuring the second half of pulse 3 followed by pulse 4. Therefore this measurement shows an additional feature of JV hysteresis at fast sweep rates due to ionic migration. We expect to be particularly sensitive to any ionic charge density at the perovskite interface due to the absence of electron/hole transporting layers in our films.
The observation of JV hysteresis at slow sweep rates in the literature could then be due to the fact that over long timescales the accumulated ions will migrate back through the film, causing a loss of the ionically charged interface and associated band-bending.

In the last part of the paper, we discuss the non-hysteretic part of the frequency-dependent conductivity in CH$_3$NH$_3$PbI$_3$, focussing particularly on the low frequency behaviour, since previous impedance spectroscopy data have shown the presence of a low frequency arc centred at \SI{0.28}{\hertz}\cite{Sanchez}/ occurring on a timescale of seconds \cite{Pascoe}. Figure~\ref{fig:3}(c) shows a simple equivalent circuit for the film, expanding on a previous model \cite{Pockett} to include the effects of ionic migration whilst neglecting any stray capacitance or substrate series resistance. $R_{\text{rec}}$ represents the electronic recombination resistance of the cell and $C_{\text{tot}}$ is the sum of the chemical and geometric capacitances, excluding any ionic effects. The additional effect of ionic migration can be modelled by a series RC circuit in parallel to the electronic processes. This physically represents an additional current channel that is blocked at the contacts, which, at constant voltage, gives an initial current of $V/R_{\text{ion}}$ that reduces exponentially to zero as $C_{\text{ion}}$ becomes fully saturated. We emphasise that simple (linear) RC circuits cannot model the hysteresis resulting from trapping behaviour, which requires nonlinear RC components \cite{Rep}. From the stored charge density of \SI{2.5}{\milli\coulomb\per\square\centi\metre} at \SI{2}{\volt} (figure \ref{fig:Fdep}), one can calculate an ionic capacitance $C_{\text{ion}}=\SI{0.28}{\milli\farad}$ , which is remarkably similar to the low frequency capacitance reported in Ref. \citenum{Sanchez}. For the parameters $R_{\text{rec}}$ and $R_{\text{ion}}$, we used an ohmic approximation to the resistance based on the current observed in the DWM at -1V (see figure~\ref{fig:Fdep}(d)), giving $R_{\text{rec}} =\SI{240}{\ohm}$, $R_{\text{ion}}=\SI{840}{\ohm}$. We approximated $C_{\text{tot}}=\SI{14}{\nano\farad}$ based on the predicted value of the relative permittivity $\epsilon_{\text{r}}$=24 in the literature \cite{Frost} and neglecting the chemical capacitance. Figure~\ref{fig:3}(d) shows the impedance response of the equivalent circuit, calculated without any adjustable parameters. 
We find that the addition of the ionic component to the equivalent circuit introduces a low frequency arc in the impedance plot, as observed in the experiments \cite{Sanchez}, but with a somewhat higher characteristic frequency $f_{\text{lf}}=\SI{0.68}{\hertz}$. It is important to stress that the exact values of the characteristic frequencies are sensitive to the geometrical parameters of the film.  For example, the relative sizes of the intermediate and low frequency arcs are determined by the ratio of $R_{\text{ion}}$ to $R_{\text{rec}}$, whilst an increase in the magnitude of the resistances will lower the characteristic frequencies and increase the width of the impedance plot (see figure~\ref{fig:3}(d)). Therefore, we conclude that our equivalent circuit model, including the effects of ionic migration, provides a good explanation of the low frequency features seen in impedance spectroscopy \cite{Sanchez, Pascoe}. Our model estimates ionic mobilities in this compound on the order of $10^{-10}$\SI{}{\square\centi\metre\per\volt\per\second}, ionic conductivities on the order of $10^{-7}$\SI{}{\per\ohm\per\centi\metre} and carrier densities on the order of $10^{20}$\SI{}{\per\cubic\centi\metre}. Further studies into the ionic mobilities of different ionic species in this compound would be beneficial in confirming this model.  We remark that, according to our explanation, the low frequency arc is purely due to conductance rather than dielectric effects \cite{dielectricnote}, as previously proposed \cite{Sanchez, Pascoe}.  Whilst some dielectric polarisation will be induced during the pulse, given the fast timescales of MA ion relaxation \cite{Poglitsch} this can only modify the JV characteristics as a function of voltage and will not contribute to the hysteresis observed.

In summary, we have completely characterised the hysteretic behaviour in CH$_3$NH$_3$PbI$_3$ thin films and have concluded that it cannot be attributed to ferroelectricity. The frequency dependent nature of the hysteretic charge is uncharacteristic of ferroelectricity, whilst its magnitude at low frequencies is over an order of magnitude larger than for well-known perovskite ferroelectrics. Importantly, we have ruled out any ferroelectric response larger than \SI{1}{\micro\coulomb\per\square\centi\metre}, significantly lower than previously suggested for this compound \cite{Frost14}. The observed hysteretic conduction effects are shown to be the result of ionic migration and of the associated charge trapping, occurring on a timescale of seconds. Consequently, it is not correct to label the observed effects in CH$_3$NH$_3$PbI$_3$ solar cells as a ferroelectric response.   Understanding the ionically induced modifications of the electrode/perovskite interface in this material is likely to be the key to explaining the JV characteristics and to enhance the performance of perovskite solar cells.

This research was funded by the Engineering and Physical Sciences Research Council grant number EP/J003557/1 and Supergen SuperSolar Project EP/J017361/1. GE thanks Oxford Photovoltaics Ltd. and the Nanotechnology KTN for funding his CASE studentship.

\bibliography{solarcellbibfile2}

\begin{thebibliography}{34}
\bibitem [{die()}]{dielectricnote}%
  \BibitemOpen
  \href@noop {} {}\bibinfo {note} {The high relative permittivity at low
  frequency must be understood in the context of lossy materials, where there
  is an additional conductive contribution to the complex relative permittivity
  $\epsilon^{'}_{\text{r}}$ = $\epsilon_{\text{r}}$ + $i \sigma/\omega$. Based
  on our value for $C_{\text{ion}}$, we obtain a low frequency value of
  $\epsilon^{'}_{r}$ of $5\times 10^{5}$, which is likely to be dominated by
  the conductive contribution.}\BibitemShut {Stop}%
\end{thebibliography}

\newpage
\includepdf{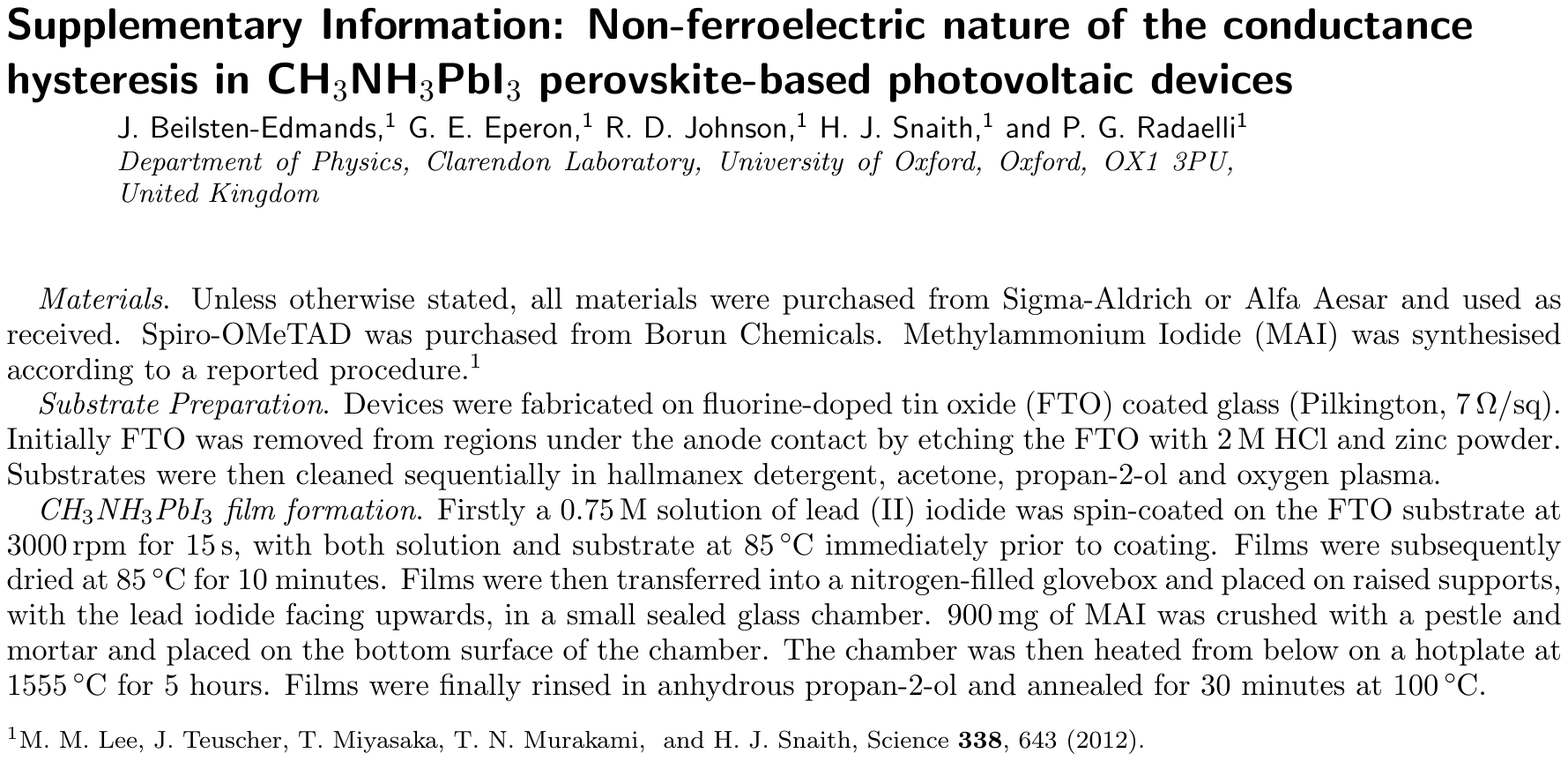}

\end{document}